\documentclass[twocolumn,showpacs,preprintnumbers,amsmath,amssymb]{revtex4}

\usepackage{graphicx}
\usepackage{dcolumn}
\usepackage{lscape}
\usepackage{bm}
\usepackage{color}

\begin{document}
\preprint{APS/123-QED}

\title{Phenomenological approach to study the degree of the itinerancy of the $5f$ electrons in actinide ferromagnets with spin fluctuation theory\footnote{This paper was presented at international conference ACTINIDE 2017 held in Sendai Japan on July 2017. It will be published in Nuclear Science and Technology.}}

\author{Naoyuki Tateiwa$^{1}$}
\email{tateiwa.naoyuki@jaea.go.jp} 
\author{Ji\v{r}{\'i} Posp{\'i}\v{s}il$^{1,2}$}
\author{Yoshinori Haga$^{1}$}%
\author{Hironori Sakai$^{1}$}%
\author{Tatsuma D. Matsuda$^{1,3}$}%
\author{Etsuji Yamamoto$^{1}$}%

\affiliation{
$^{1}$Advanced Science Research Center, Japan Atomic Energy Agency, Tokai, Naka, Ibaraki 319-1195, Japan\\
$^2$Charles University, Faculty of Mathematics and Physics, Department of Condensed Matter Physics, Ke Karlovu 5, 121 16 Prague 2, Czechia\\
$^3$Department of Physics, Tokyo Metropolitan University, Hachioji, Tokyo 192-0397, Japan\\
}
\date{\today}

\begin{abstract}
Actinide compounds with $5f$ electrons have been attracting much attention because of their interesting magnetic and electronic properties such as heavy fermion state, unconventional superconductivity, co-existence of the superconductivity and magnetism. Recently, we have reported a phenomenological analysis on 80 actinide ferromagnets with the spin fluctuation theory originally developed to explain the ferromagnetic properties of itinerant ferromagnets in the $3d$ transition metals and their intermetallics (N. Tateiwa {\it et al.}, Phys. Rev. B {\bf 96}, 035125 (2017)). Our study suggests the itinerancy of the $5f$ electrons in most of the actinide ferromagnets and the applicability of the spin fluctuation theory to actinide $5f$ system. In this paper, we present a new analysis for the spin fluctuation parameter obtained with a different theoretical formula not used in the reference. We also discuss the results of the analysis from different points of views.

 \end{abstract}

\maketitle

\section{Introduction}
The nature of the $5f$ electrons in actinide compounds has been extensively studied for many years from both theoretical and experimental points of views\cite{santini1,santini2,moore}. Despite numerous theoretical studies, the behavior of the $5f$ electrons has not been fully understood yet. One of remaining issues is whether the $5f$ electrons should be treated as being itinerant or localized in various actinide interemetallic compounds. 

Recently, we have reported the result of analyses of magnetic data in 69 uranium, 7 neptunium and 4 plutonium ferromagnets with the spin fluctuation theory developed for the magnetic properties in the $3d$ metals and their intermetallics\cite{tateiwa1}. The basic and spin fluctuation parameters of the actinide ferromagnets have been determined using our experimental data as well as those from literature. The analysis suggests the itinerancy of the $5f$ electrons in most of the actinide ferromagnets and the applicability of the spin fluctuation theory to the actinide $5f$ system. 

In this paper, we discuss the result of the analysis from different points of views. In addition, we show a new analysis for the spin fluctuation parameters obtained with a different theoretical expression in the spin fluctuation theory not used in the reference.

\section{Results and Discussions}
\subsection{Spin fluctuation theory}
We briefly summarize the spin fluctuation theory developed by Takahashi\cite{takahashi1,takahashi2,takahashi3}. The spin fluctuation spectrum for itinerant ferromagnets is described by double Lorentzian distribution functions in small energy $\omega$ and wave vector {\mbox{\boldmath $q$}} spaces.
   \begin{figure}[b]
\includegraphics[width=8.3cm]{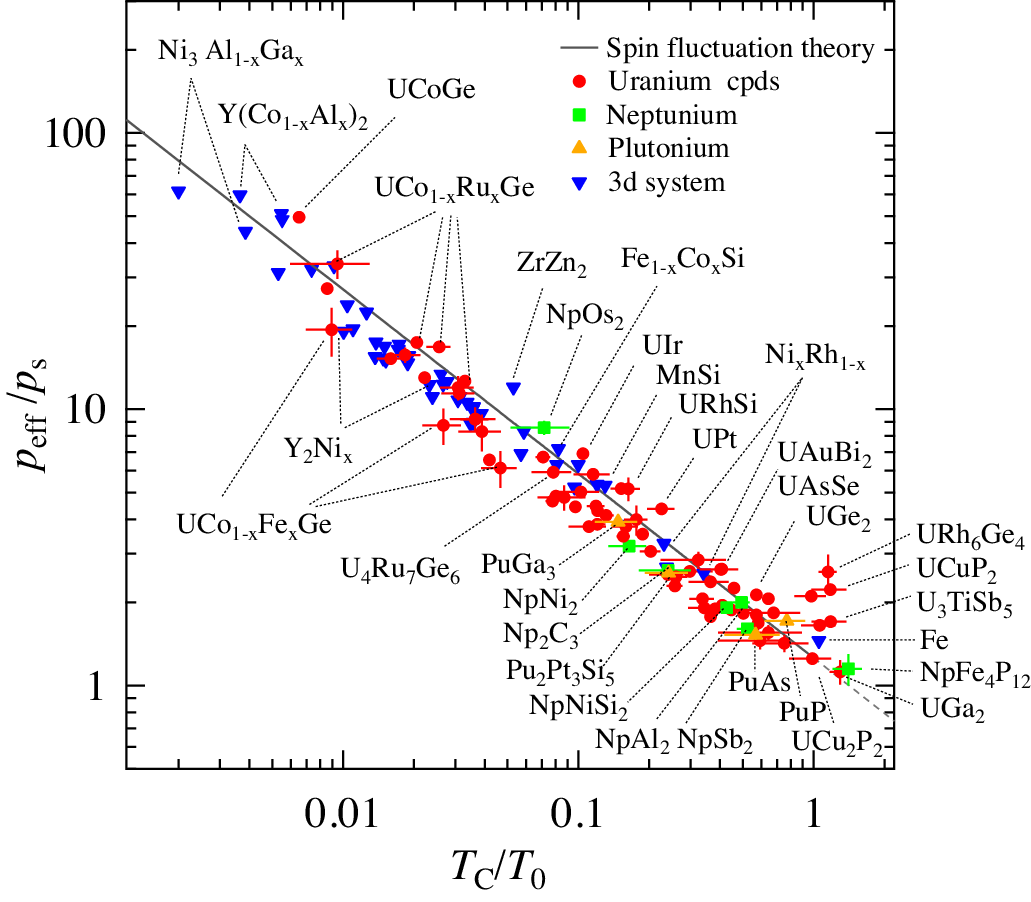}
\caption{\label{fig:epsart}Generalized Rhodes-Wohlfarth plot. Data points for uranium, neptunium and plutonium compounds, and transition $3d$ metals are plotted as closed circles, squares, triangles and anti-triangles, respectively (cited from Ref. 4). Solid line is theoretical relation between ${T_{\rm C}}/{T_0}$ and ${p_{\rm eff}}/p_{\rm s}$ in the Takahashi's spin fluctuation theory\cite{takahashi1}.}
\end{figure} 
   \begin{eqnarray}
{\rm Im} {\chi}({\mbox{\boldmath $q$}},{\omega}) &=& {{{\chi}(0)}\over 1+q^2/{\kappa}^2}{{\omega}{{\Gamma}_{q}}\over {\omega}^2+{\Gamma}_{q}^2}
 \end{eqnarray}
Here, $q{\,}{\equiv}{\,}{\lvert}{\mbox{\boldmath $q$}}{\rvert}$, ${\kappa}$ represents the inverse of the magnetic correlation length, and ${{\Gamma}_{q}}$ ($={\,}{{\Gamma}_{0}}q({\kappa}^2+q^2)$) is the damping constant. The spectrum is represented in a parameterized form by introducing two energy scales $T_0$ ($={{\Gamma}_0}q_{\rm B}^3/{2{\pi}}$) and $T_{\rm A}$ ${[={N_0}{q_{\rm B}^2}/(2{\chi}(0){{\kappa}^2}})]$, where $q_{\rm B}$ is the zone-boundary wave vector for the crystal with $N_0$ magnetic atoms with the volume $V$ ($={6{\pi}^2}{N_0}/{q_{\rm B}^3}$). These parameters represent the distribution widths of the spin fluctuation spectrum in the energy and wave-vector spaces, respectively. 

 In the Takahashi's spin fluctuation theory, $T_0$ and $T_{\rm A}$ are expressed in following relations. 
   \begin{eqnarray}
 &&{\left({{T_{\rm C}}\over{T_0}}\right)^{5/6}} = {{p_{\rm s}^2}\over {5{g^2}C_{4/3}}} {\left({15c{F_1}\over{{2}{T_{\rm C}}}}\right)^{1/2}}\\
&&  {\left({{T_{\rm C}}\over{T_{\rm A}}}\right)^{5/3}} = {{p_{\rm s}^2}\over {5{g^2}C_{4/3}}} {\left({{2}{T_{\rm C}}\over{15c{F_1}}}\right)^{1/2}}
 \end{eqnarray}
  Here, $g$ represents Lande's $g$ factor and $C_{4/3}$ is a constant ($C_{4/3}$ = 1.006089${\,}{\cdot}{\cdot}{\cdot}$). $p_{\rm s}$ is the spontaneous magnetic moment and $T_{\rm C}$ is the ferromagnetic transition temperature. $F_1$ is the mode-mode coupling term: the coefficient of the $M^4$ term in the free energy that can be evaluated experimentally from the inverse slope of the Arrott plots ($M^2$ versus $H/M$ plot) at low temperatures.  
 
  Generally, actinide ferromagnets have anisotropic magnetic properties. We analyzed the magnetic data taken on single crystal samples under magnetic field applied along the magnetic easy axis for most of the ferromagnets. But experimental data on polycrystalline samples have been used for several compounds. This could cause uncertainty in obtained parameter values that is reflected in error bars of the data points. Readers are referred to Ref. 4 for the details of our analysis.
  
  \subsection{Application of the spin fluctuation theory to actinide ferromagnets}
   \begin{figure}[b]
\includegraphics[width=8.3cm]{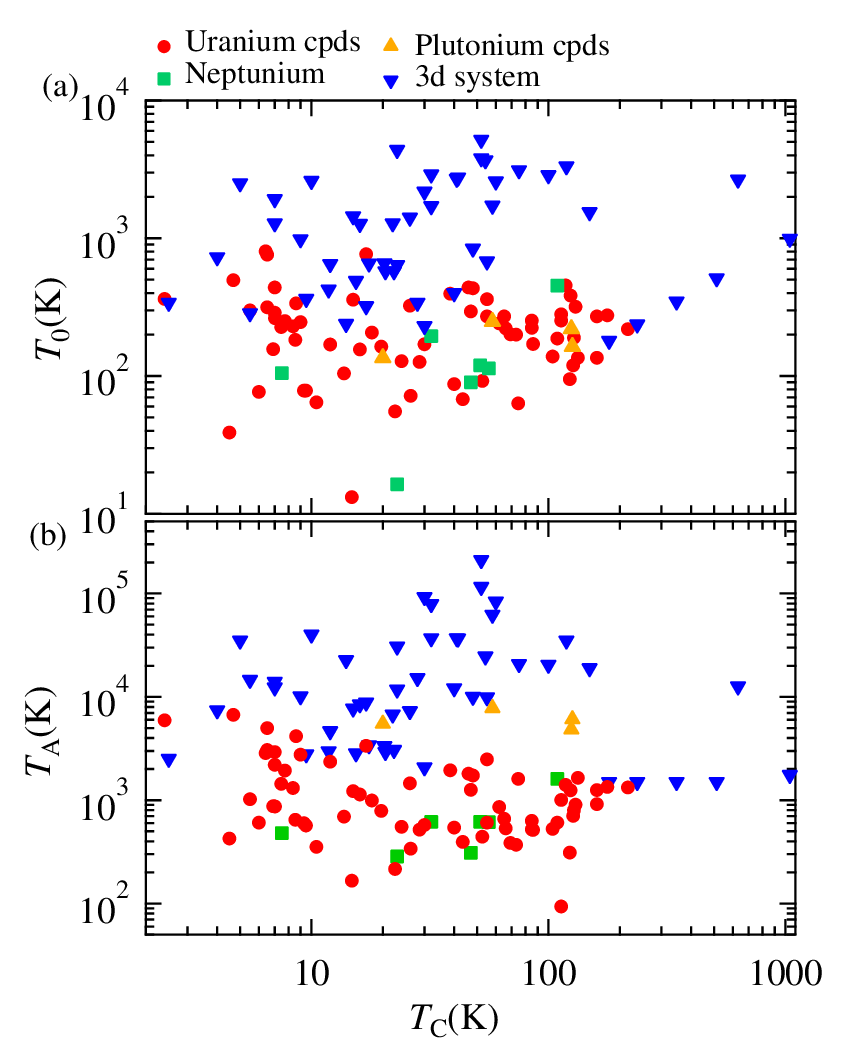}
\caption{\label{fig:epsart}Relations between $T_{\rm C}$ and (a) $T_0$ and (b) $T_{\rm A}$ for uranium, neptunium and plutonium compounds, and transition $3d$ metals plotted as closed circles, squares, triangles and anti-triangles, respectively. }
\end{figure} 

 In this subsection, we summarize our analysis on actinide ferromagnets with the spin fluctuation theory reported in Ref. 4. An important result in the Takahashi's spin fluctuation theory is generalized Rhodes-Wohlfarth relation between ${p_{\rm eff}}/{p_{\rm s}}$ and ${T_{\rm C}}/{T_0}$\cite{takahashi1,takahashi2,takahashi3}.
   \begin{eqnarray}
{p_{\rm eff}}/{p_{\rm s}}{\,}{\cong}{\,}({T_{\rm C}}/{T_0})^{-3/2}
  \end{eqnarray}
Here, $p_{\rm eff}$ is the effective magnetic moments in the magnetic susceptibility. This relation has been confirmed in itinerant ferromagnets of the $3d$ electrons systems\cite{takahashi1}. As shown in Fig. 1, we find that the same relation holds in most of actinide ferromagents for ${T_{\rm C}}/{T_0}< 1.0$\cite{tateiwa1}. This suggests the itinerant nature of the $5f$ electrons. Here, a ratio ${T_{\rm C}}/{T_0}$ characterizes the degree of itinerancy of magnetic electrons in the spin fluctuation theory. At ${T_{\rm C}}/{T_0}{\,}{\ll}{\,}1$, the magnetic electrons have a strong itinerant character. Both quantities approach unity when the degree of itinerancy becomes small. The local magnetic moment is responsible for the ferromagnetism when ${T_{\rm C}}/{T_0}{\,}={\,}1$. Note that several data points deviate from the relation for ${T_{\rm C}}/{T_0}{\,}{\sim}{\,}1.0$. This deviation may be due to several other effects not included in the spin fluctuation theory such as the crystalline electric field effect on the 5f electrons from ligand atoms. We note that the spin fluctuation theory neglects the orbital moment oriented antiparallel to the spin moment of the 5f electrons. Further elaborate theoretical consideration is necessary for the present result.

\subsection{Relation between $T_{\rm C}$ and $T_0$, and $T_{\rm A}$}
We show new result of the analysis not reported in Ref. 4. Figure 2 shows relations between $T_{\rm C}$ and (a) $T_0$ and (b) $T_{\rm A}$  for uranium, neptunium and plutonium compounds, and transition $3d$ metals plotted as closed circles, squares, triangles and anti-triangles, respectively. The parameters for the $3d$ systems are cited from Ref. 5. A general tendency is that the widths of spin fluctuation spectra in the actinide ferromagents are about one order magnitude smaller than those in the $3d$ systems. This suggests smaller energy scales of magnetic excitations for the actinide systems. This could be attributed to the larger spatial extent of the $3d$ wave functions than that of the $5f$ ones. An interesting feature is that the values of $T_{\rm A}$ for the plutonium ferromagnets are generally larger than those of the uranium and neptunium compounds and the values are comparable to those of the $3d$ systems. The spin fluctuation spectra for the plutonium ferromagnets spread to the higher momentum {\mbox{\boldmath $q$}} space, similar to the $3d$ systems.
 \begin{figure}[b]
\includegraphics[width=8.3cm]{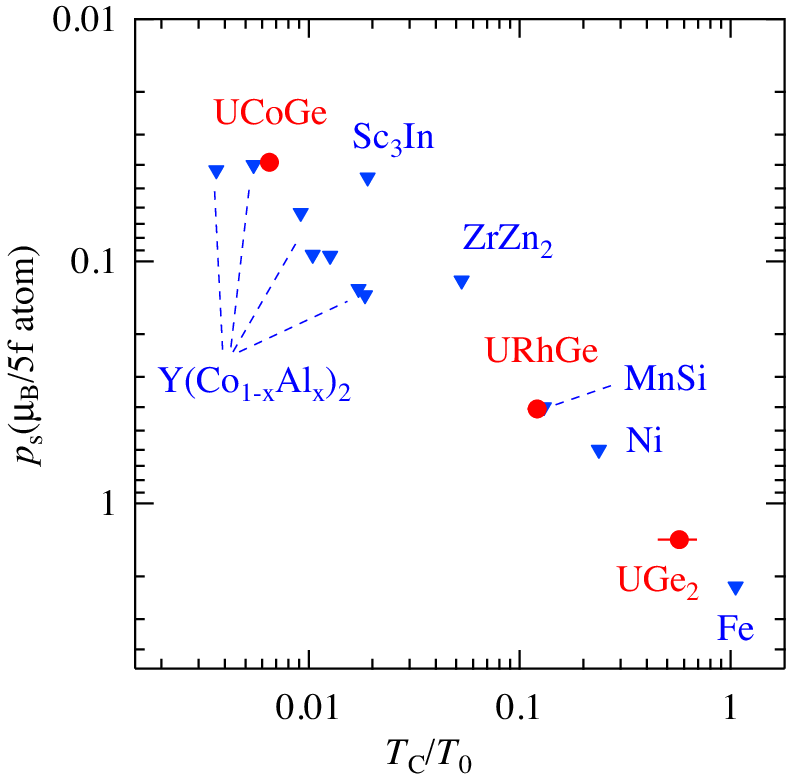}
\caption{\label{fig:epsart}Relations between ${T_{\rm C}}/{T_0}$ and $p_s$ for uranium, ferromagnetic superconductors UGe$_2$, URhGe, UCoGe, and transition $3d$ metals and their intermetallics. }
\end{figure}

\subsection{Uranium ferromagnetic superconductors UGe$_2$, URhGe, and UCoGe}
Many studies have been done for the superconducting properties in uranium ferromagnetic superconductors UGe$_2$, URhGe, and UCoGe since the findings of the superconductivity\cite{saxena,huxley0,aoki0,huy}. Meanwhile, there have been only a few studies for the dynamical magnetic property of the ferromagnetic superconductors. Figure 3 shows relations between the spontaneous magnetic moment $p_{\rm s}$ and ${T_{\rm C}}/{T_0}$ for UGe$_2$, URhGe, and UCoGe. The data points for several itinerant ferromagnets in the $3d$ system are also plotted. In this figure, right and lower region corresponds to the local moment system, and the left and upper region does the weak ferromagnetic state. The dynamical magnetic property differs depending on each ferromagnetic superconductor. The value of UGe$_2$ is 0.571, indicating that this compound is located close to the local moment system (${T_{\rm C}}/{T_0}{\,}={\,}1$). On the other hand, the value for UCoGe is 0.0065, suggesting the weak ferromagnetic state similar to those in Y(Co$_{1-x}$Al$_x$)$_2$. The values of the parameters ($T_0$ = 362 K and $T_{\rm A}$ = 5.92~103  K) in UCoGe are significantly larger than those ($T_0$ = 92.2 K and $T_{\rm A}$ = 442 K) in UGe2. The spin fluctuation spectrum ${\rm Im} {\chi}({\mbox{\boldmath $q$}},{\omega}) $ in UCoGe spreads to the higher energy and momentum spaces. URhGe is located in an intermediate region between the two limiting cases. The value of ${T_{\rm C}}/{T_0}$ for URhGe is 0.121, similar to those in MnSi (${T_{\rm C}}/{T_0}$ = 0.131) and Ni (${T_{\rm C}}/{T_0}$ = 0.237). 

  \begin{figure}[b]
\includegraphics[width=8.3cm]{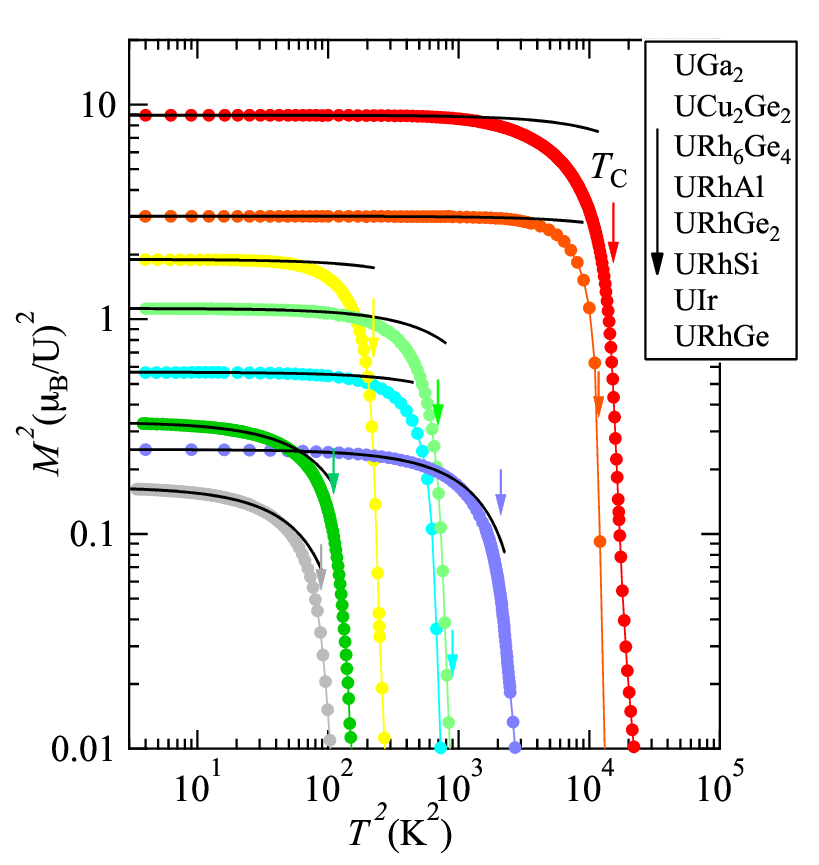}
\caption{\label{fig:epsart}Temperature dependence of the magnetization for UGa$_2$, UCu$_2$Ge$_2$, URh$_6$Ge$_4$, URhAl, URhGe$_2$, URhSi, UIr, URhGe, and URh$_{0.57}$Ir$_{0.43}$Ge under magnetic field applied along the magnetic easy axes. The Curie temperatures $T_{\rm C}$ are denoted by arrows. Solid lines are result of fits to the data with the Eq. (5).}
\end{figure} 
\begin{table}[]
\caption{\label{tab:table1}%
Table 1. Basic magnetic and spin fluctuation parameters for uranium ferromagnets classified as ggroup Ih in Ref. 4. The values of $T{_0}^0$ and $T{_{\rm A}}^0$ are cited from the reference. The values of $T{_{\rm A}}^*$ are determined using Eq. (5).  
}
\begin{ruledtabular}
\begin{tabular}{c|cc|cc|c}
&
\textrm{$T_{\rm C}$}&
\textrm{$p_{\rm s}$}&
\textrm{$T{_0}^0$}&
\textrm{$T{_{\rm A}}^0$}&
\textrm{$T{_{\rm A}}^{*}$}\\
\textrm{}&
\textrm{(K)}&
\textrm{(${\mu}_{\rm B}/{\rm U}$)}&
\textrm{(K)}&
\textrm{(K)}&
\textrm{(K)}\\
\colrule
UGe$_2$&52.6&1.41&92.2 &442 &341 \\
UGa$_2$&123 &2.94&94.8&311&348\\
UIr&46.0&0.492&440&1.80$\times$10$^3$&1.73$\times$10$^3$\\
URhGe$_2$&30.0&0.768 &170&574&612  \\
UCu$_2$Ge$_2$&109&1.74  &187&605&802\\
URhGe  &9.47 &0.407& 78.4 & 568  &531  \\
URh$_6$Ge$_4$&14.8&1.39&13.2&167&179 \\
URhAl&26.2&1.05&71.8&340&319 \\
URhSi  &10.5 &0.571&64.5&354& 323  \\
URh$_{1-x}$Co$_x$Ge &&&&&\\
$x=0.2$&13.7&0.450 &104&691& 684 \\
$x=0.6$&19.7&0.498 &164&788&726 \\
$x=0.7$&18.6&0.416&239&921&1043\\
$x=0.8$ &15.0&0.293&358&1.22$\times$10$^3$&997 \\
$x=0.9$ &7.0&0.127&439&2.19$\times$10$^3$&4.71 $\times$10$^3$\\
URh$_{1-x}$Ir$_x$Ge &&&&&\\
$x=0.15$ &9.3&0.392&78.3&599&539\\
$x=0.43$ &6.0&0.292&76.6&605&585\\
\end{tabular}
\end{ruledtabular}
 \end{table}

\subsection{Determination of  $T{_{\rm A}}^{*}$ from the $M(T)$-$T$ curve  }
 In Ref. 4, the value of the spin fluctuation parameter $T_{\rm A}$ was determined from Eqs. (2) and (3) with the mode-mode coupling term $F_1$ that is obtained from the slope of the magnetization at low temperatures\cite{tateiwa1}. Here, we estimate $T_{\rm A}$ of uranium ferromagnets classified as ``group I'' in Ref. 4 by a different theoretical formula in the Takahashi's spin fluctuation theory\cite{takahashi2,takahashi3}. The spontaneous magnetization $M(T)$ of itinerant ferromagnets is expressed by a following relation in the theory.
     \begin{eqnarray}
{\biggl(}{{M(T)}\over {M_0}}{\biggr)^2} = 1-{{a_0}\over {{p_s}^2}}{\biggl(}{{T}\over {T_{\rm A}^{*}}}{\biggr)^2}
  \end{eqnarray} 
, where $M_0$ is the spontaneous magnetization at 0 K, $a_0$ is a constant 50.4, and $p_s$ = ${M_0}/{{\mu}_{\rm B}}$. The value of $T{_{\rm A}}^*$ can be determined by the fit of the data M(T) with Eq. (5). Figure 4 shows temperature dependencies of the magnetization for UGa$_2$, UCu$_2$Ge$_2$, URh$_6$Ge$_4$, URhAl, URhGe$_2$, URhSi, UIr, and URhGe in magnetic fields of 0.01 $\sim$ 0.5 T applied along the magnetic easy axes. The Curie temperatures $T_{\rm C}$ are denoted by arrows. Solid lines are result of fits to the data with Eq. (5). Table I shows the values of $T{_{\rm A}}^*$ for the uranium ferromagnets in the group I determined in this method. The values of $T{_0}^0$ and $T{_{\rm A}}^0$ determined from the mode-mode coupling term $F_1$ using Eqs. (2) and (3) are also shown in the table. Generally, the values of $T{_{\rm A}}^0$ and $T{_{\rm A}}^*$ obtained by the two different methods are consistent with each other. This suggests the effectiveness of the theoretical expression (Eq. (5)). However, there is difference between $T{_{\rm A}}^0$ and $T{_{\rm A}}^*$ for UCu$_2$Ge$_2$, URh$_{0.2}$Co$_{0.8}$Ge, and URh$_{0.1}$Co$_{0.9}$Ge where the parameters are larger than those of the other ferromagnets. The coefficient of the $T^2$ term in Eq. (5) becomes smaller for the larger value of $T{_{\rm A}}^*$. The discrepancy may arise from ambiguity in the determination of $T{_{\rm A}}^*$ with Eq. (5). Note that the lowest temperature for the magnetic measurement is 2.0 K. The magnetization data $M(T)$ below 2.0 K may be necessary for the accurate determination of $T{_{\rm A}}^*$ when the parameter is larger.

\section{Conclusion}

We have analyzed 80 actinide ferromagnets using the spin fluctuation theory and found the itinerancy of the $5f$ electrons in most of the actinide ferromagnets. The analysis also suggests the smaller energy scales of the magnetic excitation spectra of the actinide ferromagnets. We discuss relations between the spontaneous magnetic moment $p_s$ and the spin fluctuation parameter ${T_{\rm C}}/{T_0}$ in uranium ferromagnetic superconductors UGe$_2$, URhGe, and UCoGe. We determine the parameter $T{_{\rm A}}^*$ using a different theoretical formula not used in our previous study\cite{tateiwa1}. The obtained values of $T{_{\rm A}}^*$ are basically consistent with $T{_{\rm A}}^0$ determined using the mode-mode coupling constant $F_1$. The magnetization data at very low temperatures may be necessary for the accurate determination of $T{_{\rm A}}^*$ when it is larger.

 \section{ACKNOWLEDGMENTS}
 This work was supported by Japan Society for the Promotion of Science (JSPS) KAKENHI Grant No. 16K05463, 16K05454, 16KK0106, JP15H05884(J-Physics) and 26400341.

\bibliography{apssamp}

\end{document}